\documentclass[aps,prl,twocolumn,showpacs,superscriptaddress,groupedaddress,amsmath]{revtex4}

\usepackage{graphicx}
\usepackage[usenames]{color}

\newcommand{\met}       {\mbox{$\not\!\!E_T$}}

\newcommand{\rargap}    {\mbox{ $\rightarrow$ }}
\newcommand{\ppbar}     {\mbox{$p\bar{p}$}}
\newcommand{\ttbar}     {\mbox{$t\bar{t}$}}

\newcommand{\comphep}   {\sc comphep}
\newcommand{\singletop} {\sc singletop}
\newcommand{\pythia}    {\sc pythia}
\newcommand{\alpgen}    {\sc alpgen}
\newcommand{\mcfm}      {\sc mcfm}
\newcommand{\geant}     {\sc geant}

\begin{document}

\title{Model-independent measurement of $\boldsymbol{t}$-channel single top quark production in $\boldsymbol{p\bar{p}}$ collisions at $\boldsymbol{\sqrt{s}=1.96}$~TeV }

\affiliation{Universidad de Buenos Aires, Buenos Aires, Argentina}
\affiliation{LAFEX, Centro Brasileiro de Pesquisas F{\'\i}sicas, Rio de Janeiro, Brazil}
\affiliation{Universidade do Estado do Rio de Janeiro, Rio de Janeiro, Brazil}
\affiliation{Universidade Federal do ABC, Santo Andr\'e, Brazil}
\affiliation{Instituto de F\'{\i}sica Te\'orica, Universidade Estadual Paulista, S\~ao Paulo, Brazil}
\affiliation{Simon Fraser University, Vancouver, British Columbia, and York University, Toronto, Ontario, Canada}
\affiliation{University of Science and Technology of China, Hefei, People's Republic of China}
\affiliation{Universidad de los Andes, Bogot\'{a}, Colombia}
\affiliation{Charles University, Faculty of Mathematics and Physics, Center for Particle Physics, Prague, Czech Republic}
\affiliation{Czech Technical University in Prague, Prague, Czech Republic}
\affiliation{Center for Particle Physics, Institute of Physics, Academy of Sciences of the Czech Republic, Prague, Czech Republic}
\affiliation{Universidad San Francisco de Quito, Quito, Ecuador}
\affiliation{LPC, Universit\'e Blaise Pascal, CNRS/IN2P3, Clermont, France}
\affiliation{LPSC, Universit\'e Joseph Fourier Grenoble 1, CNRS/IN2P3, Institut National Polytechnique de Grenoble, Grenoble, France}
\affiliation{CPPM, Aix-Marseille Universit\'e, CNRS/IN2P3, Marseille, France}
\affiliation{LAL, Universit\'e Paris-Sud, CNRS/IN2P3, Orsay, France}
\affiliation{LPNHE, Universit\'es Paris VI and VII, CNRS/IN2P3, Paris, France}
\affiliation{CEA, Irfu, SPP, Saclay, France}
\affiliation{IPHC, Universit\'e de Strasbourg, CNRS/IN2P3, Strasbourg, France}
\affiliation{IPNL, Universit\'e Lyon 1, CNRS/IN2P3, Villeurbanne, France and Universit\'e de Lyon, Lyon, France}
\affiliation{III. Physikalisches Institut A, RWTH Aachen University, Aachen, Germany}
\affiliation{Physikalisches Institut, Universit{\"a}t Freiburg, Freiburg, Germany}
\affiliation{II. Physikalisches Institut, Georg-August-Universit{\"a}t G\"ottingen, G\"ottingen, Germany}
\affiliation{Institut f{\"u}r Physik, Universit{\"a}t Mainz, Mainz, Germany}
\affiliation{Ludwig-Maximilians-Universit{\"a}t M{\"u}nchen, M{\"u}nchen, Germany}
\affiliation{Fachbereich Physik, Bergische Universit{\"a}t Wuppertal, Wuppertal, Germany}
\affiliation{Panjab University, Chandigarh, India}
\affiliation{Delhi University, Delhi, India}
\affiliation{Tata Institute of Fundamental Research, Mumbai, India}
\affiliation{University College Dublin, Dublin, Ireland}
\affiliation{Korea Detector Laboratory, Korea University, Seoul, Korea}
\affiliation{CINVESTAV, Mexico City, Mexico}
\affiliation{FOM-Institute NIKHEF and University of Amsterdam/NIKHEF, Amsterdam, The Netherlands}
\affiliation{Radboud University Nijmegen/NIKHEF, Nijmegen, The Netherlands}
\affiliation{Joint Institute for Nuclear Research, Dubna, Russia}
\affiliation{Institute for Theoretical and Experimental Physics, Moscow, Russia}
\affiliation{Moscow State University, Moscow, Russia}
\affiliation{Institute for High Energy Physics, Protvino, Russia}
\affiliation{Petersburg Nuclear Physics Institute, St. Petersburg, Russia}
\affiliation{Instituci\'{o} Catalana de Recerca i Estudis Avan\c{c}ats (ICREA) and Institut de F\'{i}sica d'Altes Energies (IFAE), Barcelona, Spain}
\affiliation{Stockholm University, Stockholm and Uppsala University, Uppsala, Sweden}
\affiliation{Lancaster University, Lancaster LA1 4YB, United Kingdom}
\affiliation{Imperial College London, London SW7 2AZ, United Kingdom}
\affiliation{The University of Manchester, Manchester M13 9PL, United Kingdom}
\affiliation{University of Arizona, Tucson, Arizona 85721, USA}
\affiliation{University of California Riverside, Riverside, California 92521, USA}
\affiliation{Florida State University, Tallahassee, Florida 32306, USA}
\affiliation{Fermi National Accelerator Laboratory, Batavia, Illinois 60510, USA}
\affiliation{University of Illinois at Chicago, Chicago, Illinois 60607, USA}
\affiliation{Northern Illinois University, DeKalb, Illinois 60115, USA}
\affiliation{Northwestern University, Evanston, Illinois 60208, USA}
\affiliation{Indiana University, Bloomington, Indiana 47405, USA}
\affiliation{Purdue University Calumet, Hammond, Indiana 46323, USA}
\affiliation{University of Notre Dame, Notre Dame, Indiana 46556, USA}
\affiliation{Iowa State University, Ames, Iowa 50011, USA}
\affiliation{University of Kansas, Lawrence, Kansas 66045, USA}
\affiliation{Kansas State University, Manhattan, Kansas 66506, USA}
\affiliation{Louisiana Tech University, Ruston, Louisiana 71272, USA}
\affiliation{Boston University, Boston, Massachusetts 02215, USA}
\affiliation{Northeastern University, Boston, Massachusetts 02115, USA}
\affiliation{University of Michigan, Ann Arbor, Michigan 48109, USA}
\affiliation{Michigan State University, East Lansing, Michigan 48824, USA}
\affiliation{University of Mississippi, University, Mississippi 38677, USA}
\affiliation{University of Nebraska, Lincoln, Nebraska 68588, USA}
\affiliation{Rutgers University, Piscataway, New Jersey 08855, USA}
\affiliation{Princeton University, Princeton, New Jersey 08544, USA}
\affiliation{State University of New York, Buffalo, New York 14260, USA}
\affiliation{Columbia University, New York, New York 10027, USA}
\affiliation{University of Rochester, Rochester, New York 14627, USA}
\affiliation{State University of New York, Stony Brook, New York 11794, USA}
\affiliation{Brookhaven National Laboratory, Upton, New York 11973, USA}
\affiliation{Langston University, Langston, Oklahoma 73050, USA}
\affiliation{University of Oklahoma, Norman, Oklahoma 73019, USA}
\affiliation{Oklahoma State University, Stillwater, Oklahoma 74078, USA}
\affiliation{Brown University, Providence, Rhode Island 02912, USA}
\affiliation{University of Texas, Arlington, Texas 76019, USA}
\affiliation{Southern Methodist University, Dallas, Texas 75275, USA}
\affiliation{Rice University, Houston, Texas 77005, USA}
\affiliation{University of Virginia, Charlottesville, Virginia 22901, USA}
\affiliation{University of Washington, Seattle, Washington 98195, USA}
\author{V.M.~Abazov} \affiliation{Joint Institute for Nuclear Research, Dubna, Russia}
\author{B.~Abbott} \affiliation{University of Oklahoma, Norman, Oklahoma 73019, USA}
\author{B.S.~Acharya} \affiliation{Tata Institute of Fundamental Research, Mumbai, India}
\author{M.~Adams} \affiliation{University of Illinois at Chicago, Chicago, Illinois 60607, USA}
\author{T.~Adams} \affiliation{Florida State University, Tallahassee, Florida 32306, USA}
\author{G.D.~Alexeev} \affiliation{Joint Institute for Nuclear Research, Dubna, Russia}
\author{G.~Alkhazov} \affiliation{Petersburg Nuclear Physics Institute, St. Petersburg, Russia}
\author{A.~Alton$^{a}$} \affiliation{University of Michigan, Ann Arbor, Michigan 48109, USA}
\author{G.~Alverson} \affiliation{Northeastern University, Boston, Massachusetts 02115, USA}
\author{G.A.~Alves} \affiliation{LAFEX, Centro Brasileiro de Pesquisas F{\'\i}sicas, Rio de Janeiro, Brazil}
% ...
%
%
\collaboration{The D0 Collaboration\footnote{with visitors from
$^{a}$Augustana College, Sioux Falls, SD, USA,
$^{b}$The University of Liverpool, Liverpool, UK,
$^{c}$SLAC, Menlo Park, CA, USA,
$^{d}$University College London, London, UK,
$^{e}$Centro de Investigacion en Computacion - IPN, Mexico City, Mexico,
$^{f}$ECFM, Universidad Autonoma de Sinaloa, Culiac\'an, Mexico,
and 
$^{g}$Universit{\"a}t Bern, Bern, Switzerland.
$^{h}$Visitor from ETH-Zurich
}} \noaffiliation
\vskip 0.25cm

\date{May 13, 2011}

\begin{abstract}
We present a model-independent measurement of $t$-channel electroweak production of single top quarks
in $\ppbar$ collisions at $\sqrt{s}=1.96\;\rm TeV$. Using $5.4\;\rm fb^{-1}$
of integrated luminosity collected by the D0 detector at the Fermilab Tevatron Collider, 
and selecting events containing 
an isolated electron or muon, missing transverse energy and one or two jets
originating from the fragmentation of $b$ quarks, we measure a cross section 
$\sigma({\ppbar}{\rargap}tqb+X) = 2.90 \pm 0.59\;\rm (stat+syst)\; pb$ for a top quark mass 
of $172.5\;\rm GeV$. The probability of the background to fluctuate and 
produce a signal as large as the one observed is $1.6\times10^{-8}$, corresponding to
a significance of $5.5$ standard deviations.
\end{abstract}


\maketitle 




Within the standard model (SM), top quarks are produced through two 
different types of interactions at hadron colliders. 
The largest yield is from \ttbar\ pairs, produced via the strong interaction~\cite{ttbar-xsec}. 
Pair production was used by the D0 and CDF 
collaborations to establish the existence of the top quark in 
1995~\cite{top-obs-1995-d0,top-obs-1995-cdf}. 
The second source of top quarks is the 
electroweak production of single top quarks, observed 
by the D0 and CDF collaborations in 2009~\cite{stop-obs-2009-d0,stop-obs-2009-cdf}.
The single top quark production rate is about half that for \ttbar\ pairs 
and the signal-to-background ratio is worse, so the observation of this process is more difficult.  

At the Tevatron, single top quarks are produced through $W$ boson exchange, and accompanied
by a $b$ quark in the 
$s$ channel ($t\bar{b}+\bar{t}b = tb$)~\cite{singletop-cortese}, 
or by both a $b$ and a light quark in the $t$-channel 
($tq\bar{b}+\bar{t}qb = tqb$)~\cite{singletop-willenbrock,singletop-yuan}.
A third process, usually called ``associated production,'' 
in which the top quark is produced together with a $W$ boson, has negligible 
cross section at the Tevatron~\cite{singletop-xsec-kidonakis}.
Prior D0 and CDF publications~\cite{stop-obs-2009-d0,stop-obs-2009-cdf} measured the production cross section for
the sum of $tb$ and $tqb$ while 
assuming the relative rate between the two processes as predicted 
by the SM. This relative rate, however, departs from the SM prediction in several
new physics scenarios, for example, in models with additional quark generations~\cite{Alwall:2007},
new heavy bosons~\cite{Tait:2000sh}, flavor-changing neutral currents (FCNC)~\cite{d0-fcnc}, or
anomalous top quark couplings~\cite{singletop-wtb-heinson,d0-singletop-wtb,Abazov:2009ky}.
It is therefore of interest to measure the $tb$ and $tqb$ production cross sections separately, 
using a methodology that is independent of theoretical assumptions about their relative rates~\cite{Tait:2000sh}. 
Furthermore, 
measuring the production rates of the two single top quark production modes to a precision 
comparable to that of the \ttbar\ pair cross section not only tests the validity of the 
parton distribution functions of the proton and antiproton, 
but also the Cabibbo-Kobayashi-Maskawa (CKM) matrix elements~\cite{CKM} 
and higher-order corrections from quantum chromodynamics. 
These high-precision comparisons are important to check the self-consistency of the standard model (SM). 

This Letter builds on the techniques developed by the 
D0 collaboration for prior measurements of single top
quark production~\cite{d0-prl-2007, d0-prd-2008, stop-obs-2009-d0, $t$-channel},
and analyzes a larger dataset corresponding to an integrated luminosity 
of $5.4\;\rm fb^{-1}$. Using improved multivariate 
analysis (MVA) techniques, 
we report a new measurement of the $tqb$ production rate that makes no assumptions about the $tb$ 
production rate, and is therefore sensitive to various new physics scenarios.


The data for this analysis were collected with a logical OR of many trigger 
conditions that results in a fully efficient trigger selection for the single top quark signal.  
Electron, muon, missing transverse energy ($\met$), and jet reconstruction algorithms used in this analysis 
have been described elsewhere~\cite{d0-prd-2008, d0-detector}. 
Events are selected that contain  
one jet with transverse momentum $p_T>25\;\rm GeV$ and at least a second jet with $p_T>15\;\rm GeV$, 
both within pseudorapidity $|\eta|<3.4$.
Events with 2, 3 or 4 jets are used in the analysis. 
Events are also required to contain exactly one isolated high-$p_T$ electron or muon that originates from 
the $\ppbar$ interaction vertex and satisfies the following acceptance criteria:
for the electron $|\eta| < 1.1$ and $p_T > 15 (20)\;\rm GeV$ for events with 2 (3 or 4) jets; 
for the muon $|\eta| < 2.0$ and $p_T > 15\;\rm GeV$.  
The $\met$ is required to be in the range of $(20,200)\;\rm GeV$ for events with 2 jets and 
$(25,200)\;\rm GeV$ for events with 3 or 4 jets. 

The main background contributions arise from $W$~bosons
produced in association with jets, {\ttbar} pairs, and multijet
production in which a jet with high electromagnetic
content mimics an electron, or a muon contained within a jet originating from
the decay of a heavy-flavor quark ($b$ or $c$ quark) appears isolated.
Diboson ($WW$, $WZ$, $ZZ$) and  $Z$+jets processes 
add small additional contributions to the background. 
To limit the number of events originating from multijet production that enter our candidate samples
through $\met$ mismeasurements,  
we require the $\met$ to be not aligned with or opposing the direction of the lepton or the leading jet 
and require high total scalar transverse energy of the lepton, jets and $\met$ ($H_T$).
Additional requirements are placed on the transverse energy of all jets in all events
and the second leading jet $p_T$ in two-jet events to suppress soft-scattering background.

To enhance the fraction of signal events in data, one or two of the jets in each event 
are required to originate from the hadronization of long-lived $b$~hadrons, as 
inferred from a neural network $b$-tagging algorithm ("b-tagged jet")~\cite{D0btag}. 
The data are divided into six mutually-exclusive subsamples
to take advantage of different signal-to-background ratios and different sources 
of background. This separation is based on jet multiplicity (2, 3, or 4 jets), and
on the number of b-tagged jets (1 or 2 $b$-tags).  
The signal acceptance of all selections corresponds to  $(2.0 \pm 0.3)\%$ for $tqb$ and
$(2.9 \pm 0.4)\%$ for $tb$ production, as estimated using Monte Carlo (MC) simulations.


Single top quark signal events are modeled 
using the {\comphep}-based MC event generator
{\singletop}~\cite{singletop-mcgen}. 
The kinematics of the generated events
closely match those predicted by next-to-leading-order
(NLO) calculations~\cite{sullivan}.
{\pythia}~\cite{pythia} is used to model the
hadronization of all generated partons. 
The {\ttbar}, $W$+jets, and $Z$+jets backgrounds are simulated using
the {\alpgen} leading-log MC event generator~\cite{alpgen}, with 
{\pythia} used to model hadronization. The {\ttbar} background is
normalized to the predicted cross section~\cite{ttbar-xsec} for a top quark mass of $172.5\;\rm GeV$. 
The normalization of the $W/Z$ backgrounds is obtained by scaling the {\alpgen} cross sections 
to account for NLO effects~\cite{mcfm}. Diboson backgrounds are modeled using {\pythia}.

The presence of additional $\ppbar$ 
interactions is modeled by events selected from random beam crossings matching the 
instantaneous luminosity profile in the data.
All MC events are passed through a {\geant}-based
simulation~\cite{geant} of the D0 detector. 
Small corrections, derived using samples depleted of single top quark events, are applied to all reconstructed objects to improve the agreement between data and simulation.
In particular, any mismodeling of the pseudorapidity of jets 
or of the angular separation between the two leading jets in the $W$+jets
sample is adjusted by reweighting the MC.

The multijet background is modeled using data 
containing non-isolated leptons. 
In the electron channel, the transverse momentum distribution of the lepton is reweighted to match 
the distribution in background events that pass the candidate-selection criteria. 
To increase the statistics in the muon channel, the jet
closest to the muon is removed and the {\met} recalculated. 
The overall normalization of the $W$+jets and the multijet backgrounds is obtained 
by comparing the background expectation to data in three sensitive variables:
lepton transverse momentum, {\met}, and the transverse mass of the $W$~boson. This normalization is found  
after subtracting the contributions from the small backgrounds 
(\ttbar, $Z$+jets, and dibosons), before $b$-tagging, when the expected signal-to-background
ratio (S:B) for the entire sample is S:B $\approx$ 1:279. 

The probability for the $b$-tagging algorithm to identify a jet as originating from a $b$ quark is 
measured in a sample of data with jets containing muons and compared to the corresponding 
efficiency measured in a MC sample. From this 
comparison we derive correction factors that we parametrize as a function of
jet flavor, $p_T$, and $\eta$ and apply to all the MC samples used in this analysis.
After $b$-tagging, we check the normalization of the $Wb\bar{b}$ and $Wc\bar{c}$ samples in 
a sample that has no overlap with the one used in the single top quark cross section measurement and 
find the normalization to be consistent with unity. We assign a 12\% uncertainty 
on the normalization of the $Wb\bar{b}$ and $Wc\bar{c}$ samples
to account for variations in the 
$Wcj$ ($j=u,d,s$) cross section and the $Wb\bar{b}$ to $Wc\bar{c}$ cross section ratio used in this study.

We consider systematic uncertainties that arise from the theoretical and detector modeling of the background processes 
and from the subsequent corrections applied to that model. Most uncertainties affect only the normalization 
of the distributions, 
but the corrections for the jet energy scale (JES), the jet energy resolution (JER), and the 
reweighting of the kinematic distributions in $W$+jets events also modify the differential distributions. 
The largest uncertainties come from the JER, corrections to the $b$-tagging efficiency, and
the corrections for the jet-flavor composition in $W$+jets events, with smaller contribution from JES, 
MC statistics, integrated luminosity, and trigger uncertainties. The total systematic
uncertainty on the background is 11\%.

We select 8,471 events and expect $238\pm 28$ $tqb$ events.
This corresponds to a S:B of approximately 1:33 for the sample with 1 $b$-tag and 
1:50 for the sample with 2 $b$-tags.
Table~\ref{tab:event-yields} lists the number of events expected and observed for each process as a
function of jet multiplicity.    
We also define two control samples to check the background model components 
separately for the two main backgrounds: $W$+jets and \ttbar. The $W$+jets-dominated sample has low $H_T$, 
exactly two jets,  
and only one $b$-tagged jet. The \ttbar-dominated sample has high $H_T$, 
exactly four jets, and one or two $b$-tagged jets. 
We find good agreement for both normalization and shape in most of distributions studied 
for the cross-check samples and also for the candidate sample before $b$-tagging, when the signal 
contribution is negligible. 

\begin{table}[!h!tbp]
\vspace{-0.15in}
\begin{center}
\caption{Number of expected and observed events in $5.4\;\rm fb^{-1}$ of integrated luminosity,
with uncertainties including both statistical and systematic components. The $tb$ and $tqb$ contributions are
normalized to their SM expectations.
}
\label{tab:event-yields}
\begin{ruledtabular}    
\begin{tabular}{lccc}   
Source     &     2 jets      &     3 jets     &     4 jets     \\
\hline
$tb$       & 104 $\pm$ 16 & 44 $\pm$ 7.8 & 13 $\pm$ 3.5 \\
$tqb$      & 140 $\pm$ 13 & 72 $\pm$ 9.4 & 26 $\pm$ 6.4 \\
${\ttbar}$ & 433 $\pm$ 87 & 830 $\pm$ 133 & 860 $\pm$ 163 \\
$W$+jets   & 3,560 $\pm$ 354 & 1,099 $\pm$ 169 & 284 $\pm$ 76 \\
$Z$+jets \& dibosons & 400 $\pm$ 55 & 142 $\pm$ 41 & 35 $\pm$ 18 \\
Multijets  & 277 $\pm$ 34 & 130 $\pm$ 17 & 43 $\pm$ 5.2 \\
Total prediction & 4,914 $\pm$ 558 & 2,317 $\pm$ 377 & 1,261 $\pm$ 272 \\
Data       & 4,881 & 2,307 & 1,283
\end{tabular}
\end{ruledtabular}
\end{center}
\vspace{-0.15in}
\end{table}


We further improve the discrimination between signal and background by 
employing MVA techniques as described in~\cite{d0-prd-2008}. 
We use two of our previous methods: 
boosted decision trees (BDT)~\cite{decision-trees} and Bayesian neural
networks (BNN)~\cite{bayesianNNs}. In addition, we use a third method, 
neuroevolution of augmented topologies (NEAT)~\cite{neat}. 
All three methods use the same data and background model, consider the same
sources of systematic uncertainty, and rely on 
variables with signal-background discrimination power chosen from a common set 
of well modeled variables from five categories~\cite{variables-schwienhorst,Cao:2005pq}: 
single object kinematics, global event kinematics, jet 
reconstruction, top quark reconstruction, and angular correlations. 
The BDT method uses a common set of 
50 variables for all analysis channels, while the NEAT method
uses the TMVA implementation of the ``RuleFit'' algorithm~\cite{rulefit,tmva} to select the 30 most
sensitive kinematic variables in each channel. New for this analysis, the BNN method 
constructs a discriminant using the lepton and jet 4-vectors, the $\met$ 2-vector, and variables 
that include lepton charge and $b$-tagging information, for a total of 14, 18 and 22 variables for events with 2, 3 and 4 jets.

\begin{figure}[!h!tbp]
\centering
\includegraphics[width=0.238\textwidth]{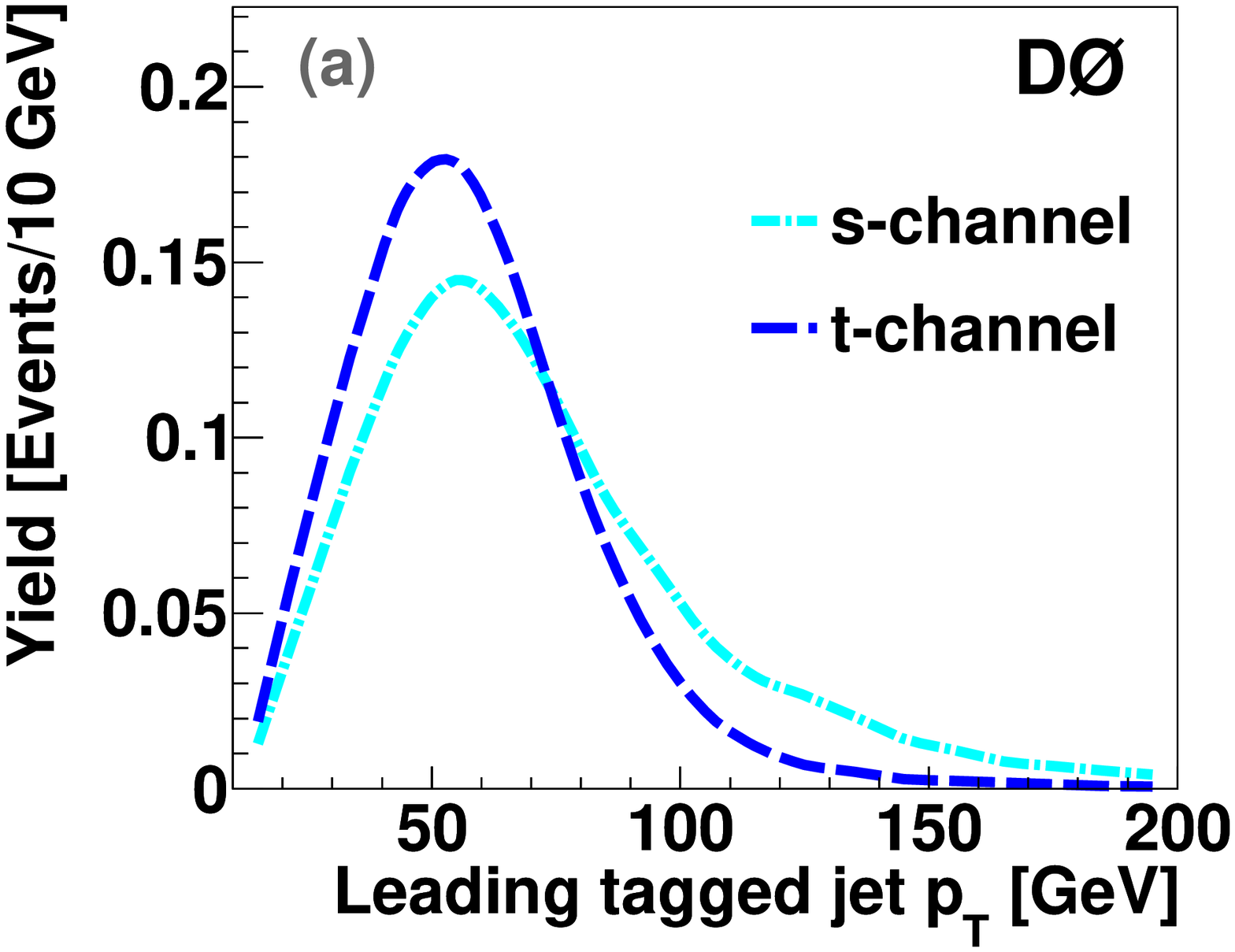}
\includegraphics[width=0.238\textwidth]{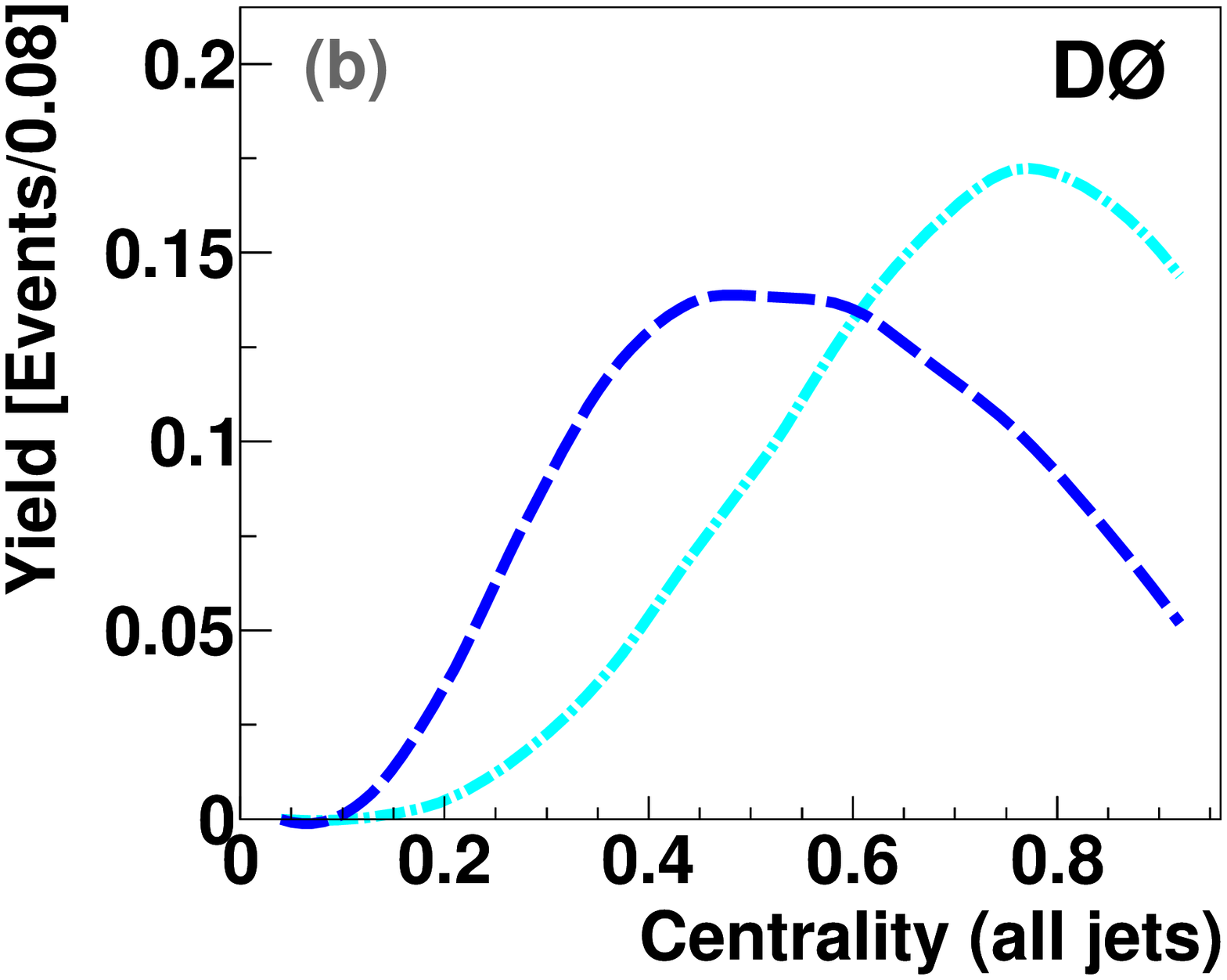}
\includegraphics[width=0.238\textwidth]{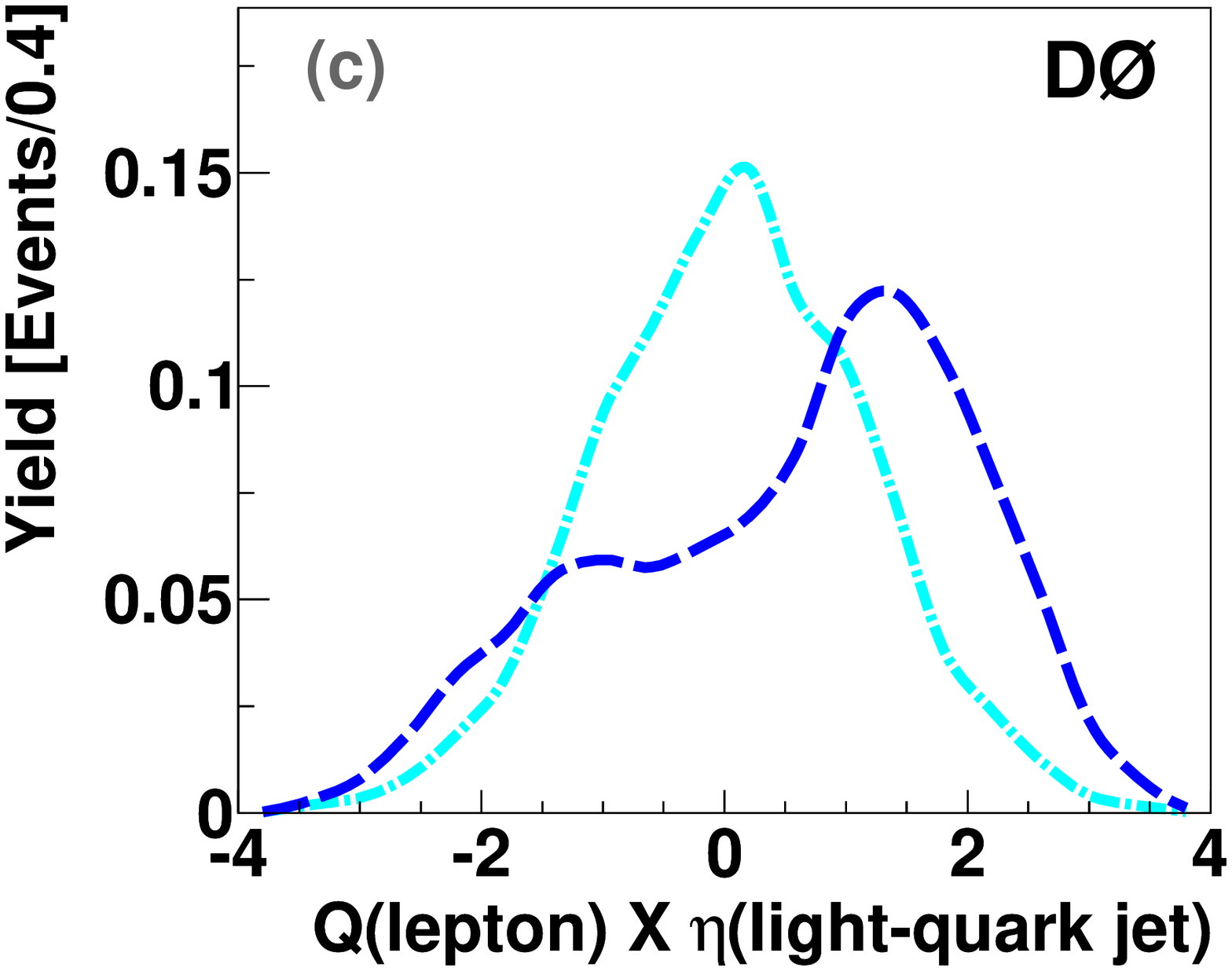}
\includegraphics[width=0.238\textwidth]{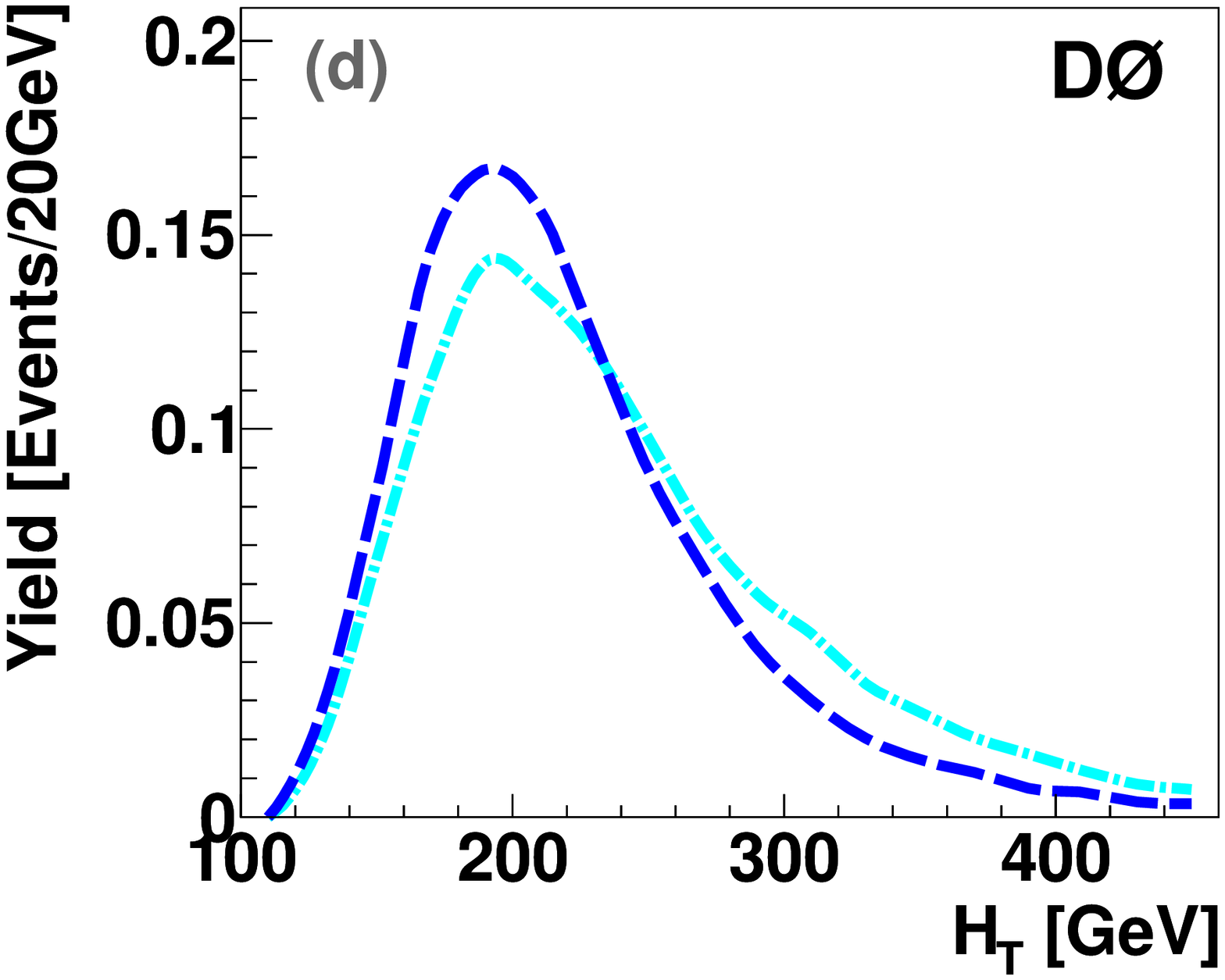}
\vspace{-0.2in}
\caption{Distribution for $tb$ and $tqb$ processes for (a) transverse momentum of the leading $b$-tagged jet, 
(b) centrality, defined as
the ratio of the scalar sum of the $p_T$ of the jets to the scalar
sum of the energy of the jets in the event, 
(c) lepton charge multiplied by the leading non-$b$-tagged quark jet pseudorapidity and 
(d) total transverse energy. All the distributions are normalized to unity.
}
\label{fig:norm}
\end{figure}

Each method is optimized to maximize the sensitivity to the $tqb$ signal 
by treating the $tb$ process as a background component with normalization given by the SM cross section~\cite{singletop-xsec-kidonakis}. 
Figure~\ref{fig:norm} 
shows the distributions of some of the variables that provide the greatest discrimination 
between the $tqb$ and $tb$ processes.
To further improve the sensitivity, the analyses are 
performed separately on the six mutually exclusive subsamples defined before. 
For all three methods the output variables are saved in histograms with a 
binning chosen to ensure that there are enough events to limit 
the uncertainties due to MC statistics.

\begin{figure}[!h!tbp]
\centering
\includegraphics[width=0.238\textwidth]{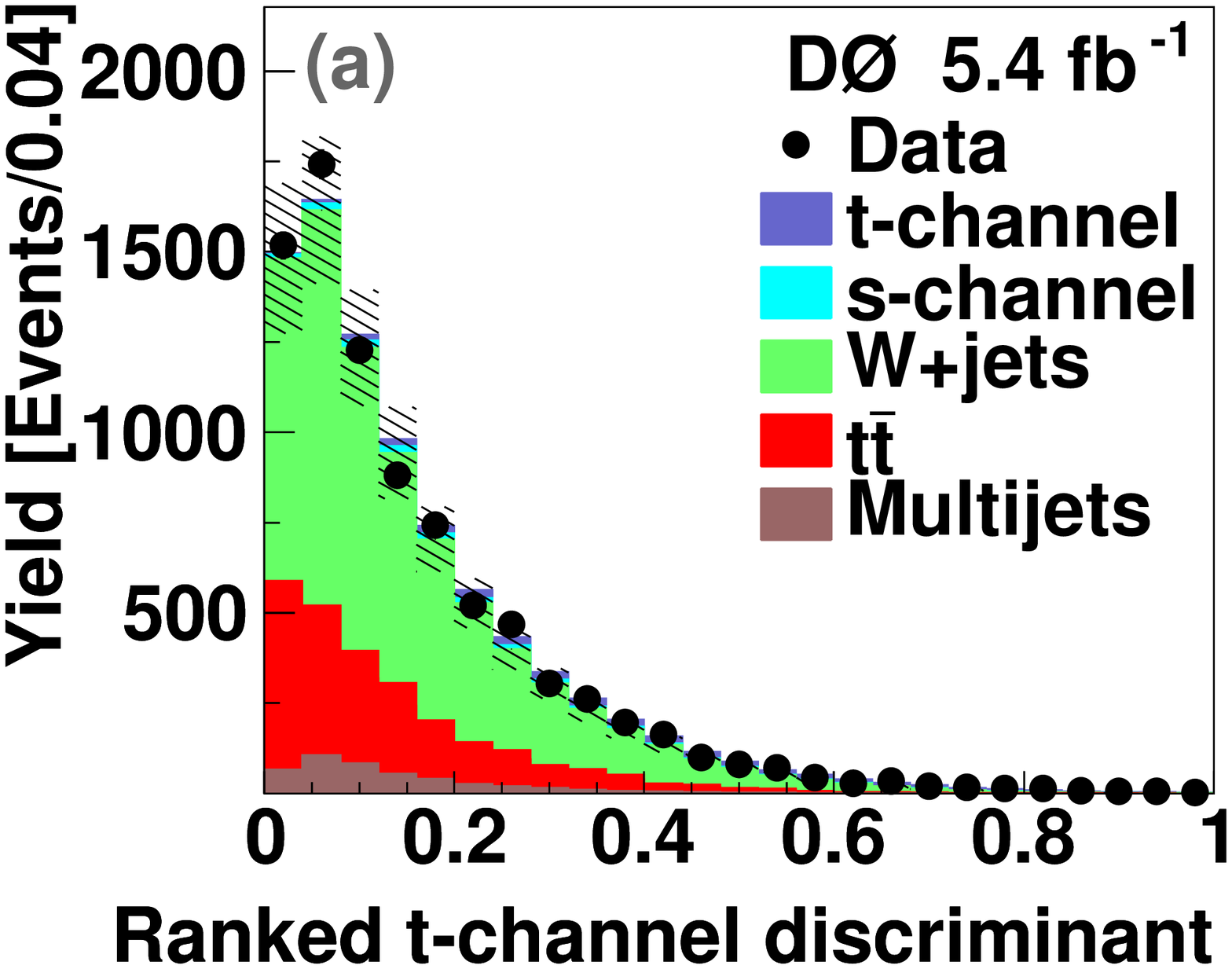}
\includegraphics[width=0.238\textwidth]{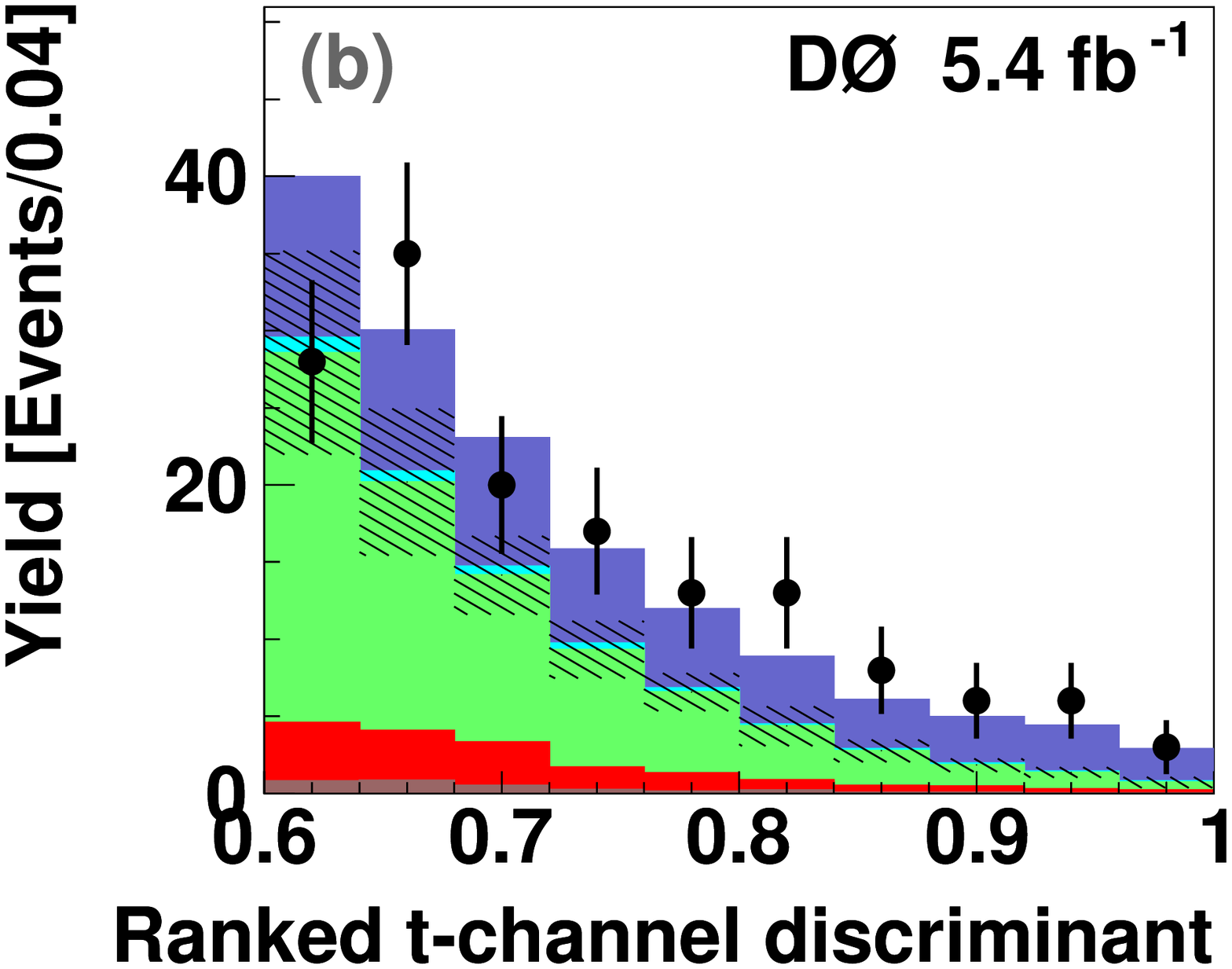}
\vspace{-0.2in}
\caption{Comparison of the signal and background models to data for the combined $t$-channel discriminant
for (a) the entire discriminant range and (b) the signal region.
The bins have been ordered by their expected S:B.
The single top quark contributions are normalized to the measured cross sections. The $t$-channel contribution 
is visible above the hatched bands that show the uncertainty on the background prediction.
}
\label{fig:BnnCombT}
\end{figure}

Even though the three MVA techniques use the same data sample, they are only 
$\approx 70\%$ correlated with each other. We therefore combine these methods using 
an additional BNN algorithm (BNNComb) that takes
as input the individual output discriminants of the BDT, BNN, and NEAT methods, 
and produces a single combined output discriminant. 
Figure~\ref{fig:BnnCombT} shows comparisons between the $t$-channel signal,
the background model, and data for the combined discriminant,
which leads to a more precise measurement of the cross section.

The single top quark production cross section is measured using a Bayesian 
approach as in~\cite{d0-prl-2007, d0-prd-2008, stop-obs-2009-d0}. 
We follow the approach of \cite{$t$-channel} and construct a two-dimensional (2D)
posterior probability density as a 
function of the cross sections for the $tqb$ and $tb$ processes. 
The output discriminants for the signals, backgrounds, and data 
are used to form a binned likelihood as a
product over all six analysis channels and all bins. 
No constraint is imposed on the relative rates of $tb$ and $tqb$ 
production. We assume a Poisson distribution for the observed number of events and uniform 
prior probabilities with positive values for the two signal cross sections. 
We integrate over the systematic uncertainties which are described by Gaussian
priors that preserve the correlations between bins and channels. 
The $tqb$ cross section is then extracted from a one-dimensional posterior
probability density obtained from this 2D posterior 
by integrating over the $tb$ axis, thus not making any assumptions about
the value of the $s$-channel cross section. 
Similarly, the $tb$ cross section is obtained by integrating over the $tqb$ axis. 
Ensembles of datasets generated at several different cross section values are used to verify the linearity of the cross section extraction procedure.

\begin{figure}[!h!tbp]
\vspace{-0.1in}
\centering
\includegraphics[width=3.0in]{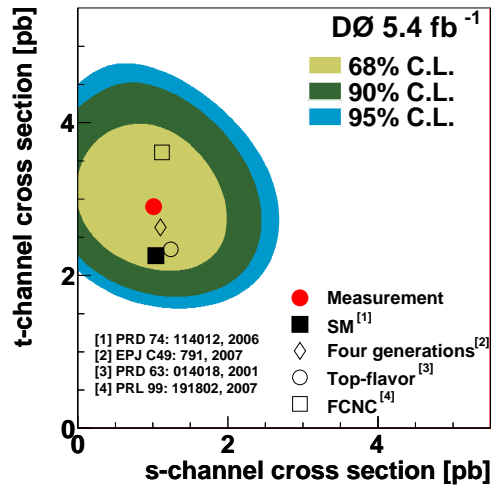}
\vspace{-0.2in}\\
\caption{Posterior probability density for $tqb$ vs $tb$ single top quark
production in contours of equal probability density. The measured
cross section and various theoretical predictions are also shown.}
\label{fig:2dposterior}
\end{figure}

Figure~\ref{fig:2dposterior} shows the 
2D posterior probability density for the combined discriminant 
together with predictions from the SM~\cite{singletop-xsec-kidonakis} and various beyond-the-SM scenarios: 
four-quark-generations with CKM matrix element $|V_{ts}|=0.2$~\cite{Alwall:2007}, 
top-flavor model with new heavy bosons at a scale $m_x=1$~TeV~\cite{Tait:2000sh}, 
and FCNC with an up-quark/top-quark/gluon coupling
$\kappa_u/\Lambda=0.036$~\cite{d0-fcnc}.

The measured cross sections of $\sigma({\ppbar}{\rargap}tqb+X) = 2.90 \pm 0.59$ pb and 
$\sigma({\ppbar}{\rargap}tb+X) = 0.98 \pm 0.63$ pb are in good agreement with the 
SM expectation for a top quark mass of $172.5\;\rm GeV$~\cite{singletop-xsec-kidonakis}.
The uncertainty includes both statistical and systematic sources.
The cross section for $t$-channel single top quark production is the most precise measurement of an 
individual single top quark production channel to date with an uncertainty of 20\%.

The significance of the $t$-channel cross section
measurement is computed using a log-likelihood ratio
approach~\cite{stop-obs-2009-cdf, $t$-channel} which tests
the compatibility of the data with two hypotheses: a null hypothesis
where there is only background and a background plus signal hypothesis, where the number of signal events corresponds to
the theoretical cross section. New for this analysis is the computation of the distributions 
for these two hypotheses given by an asymptotic Gaussian approximation~\cite{ref:fast-significance}. 
With this approximation we compute for the first time, the significance of the 
measured $tqb$ cross section independently of any assumption on the production rate of $tb$. 
We estimate the probability of the background to fluctuate and 
produce a signal as large as the one observed to be $1.6\times10^{-8}$, corresponding to
a significance of $5.5$ standard deviations (SD). The expected significance is 4.6 SD.

\begin{figure}[!h!btp]
\centering
\vspace{-0.1in}
\hspace{-0.1in}
\includegraphics[width=0.246\textwidth]{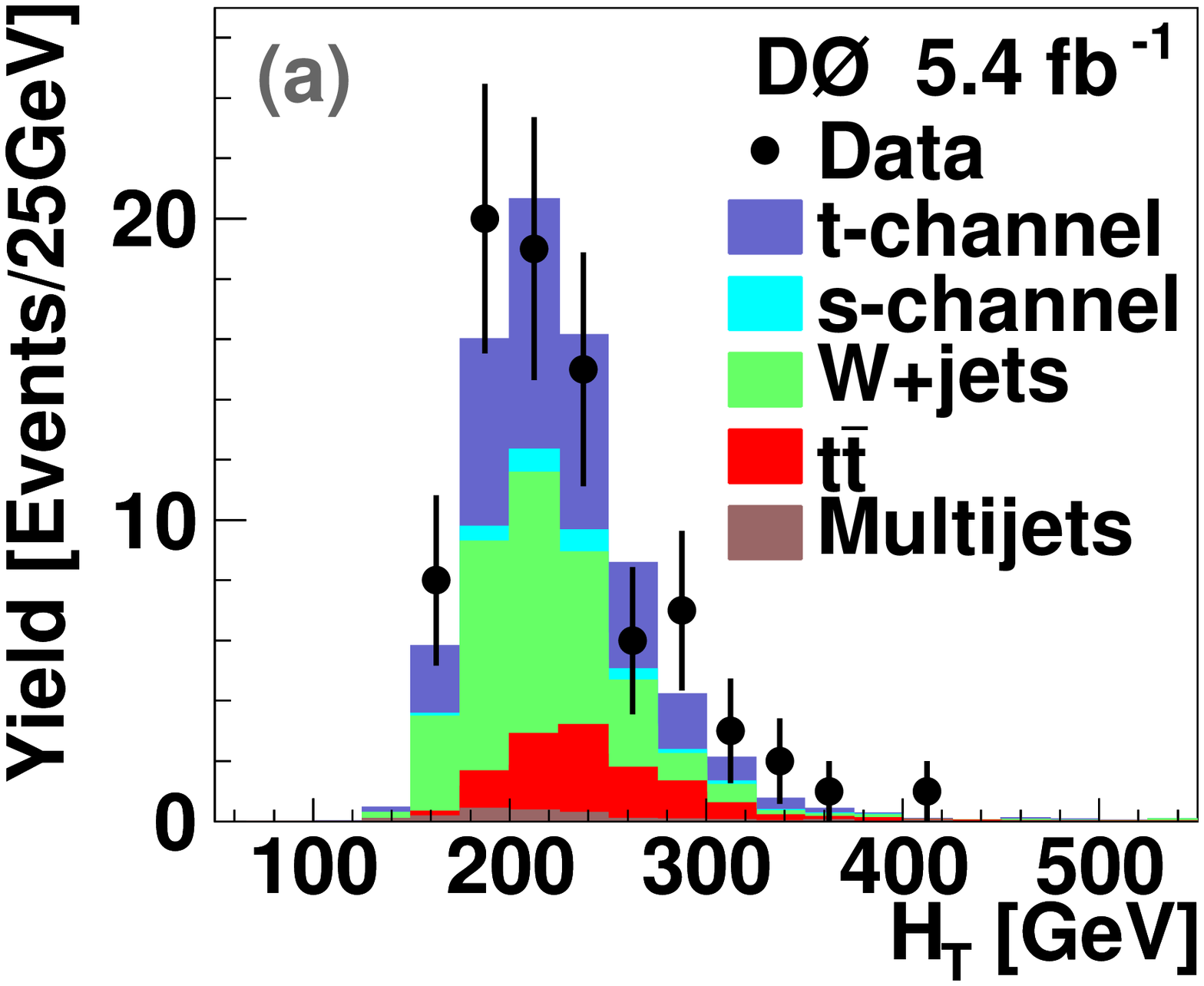}
\hspace{-0.1in}
\includegraphics[width=0.246\textwidth]{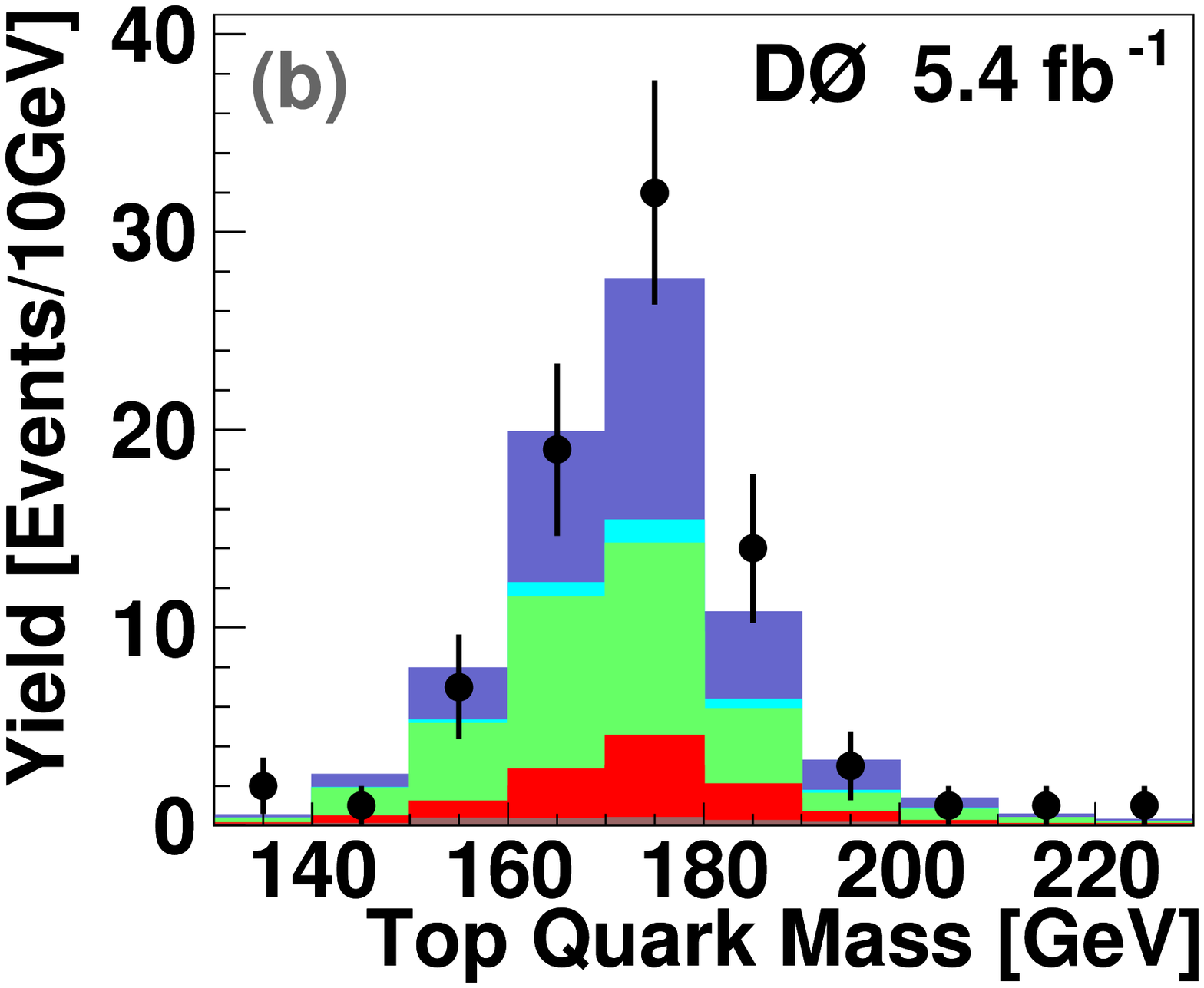} 
\\ \hspace{-0.1in}
\includegraphics[width=0.246\textwidth]{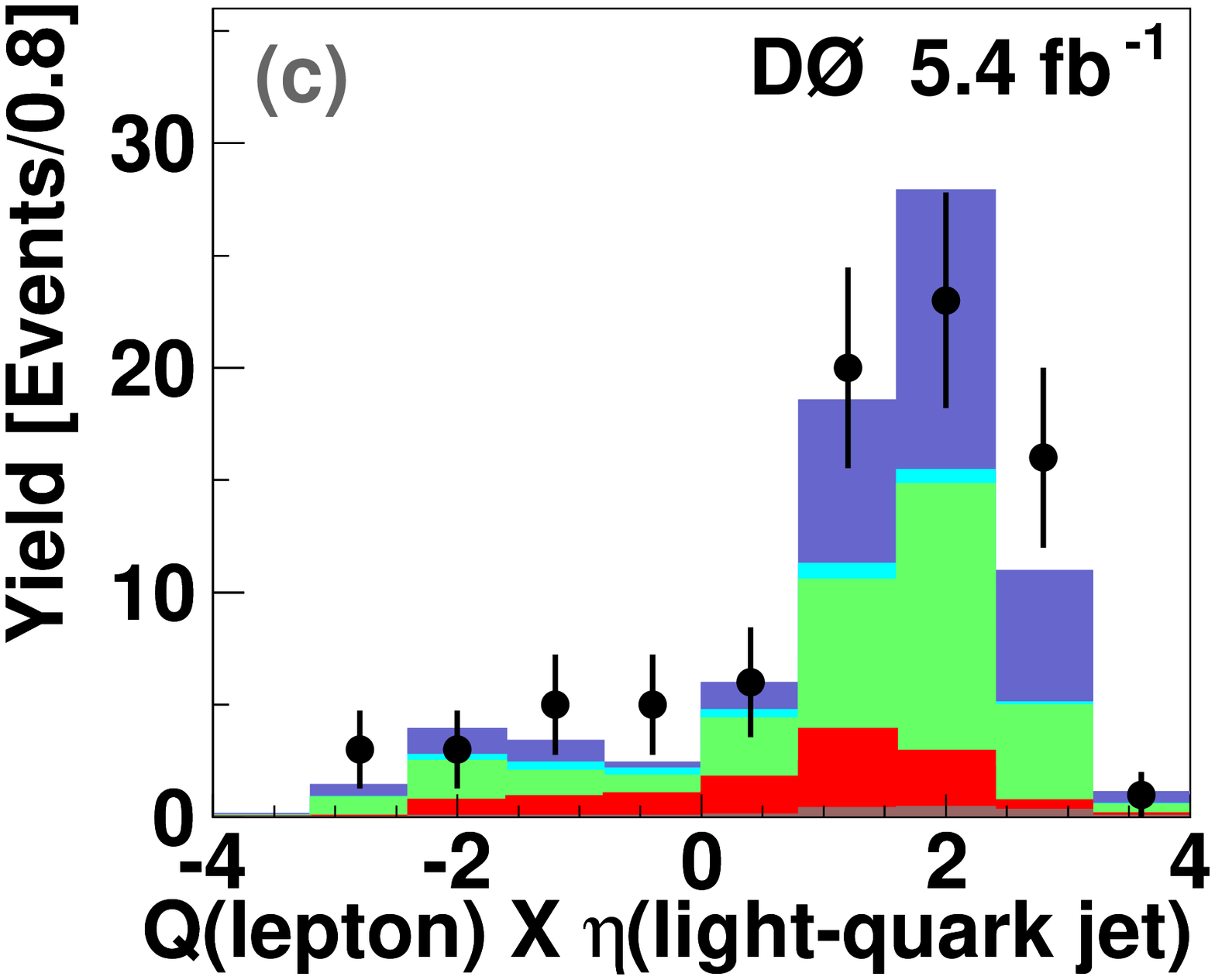}
\hspace{-0.1in}
\includegraphics[width=0.246\textwidth]{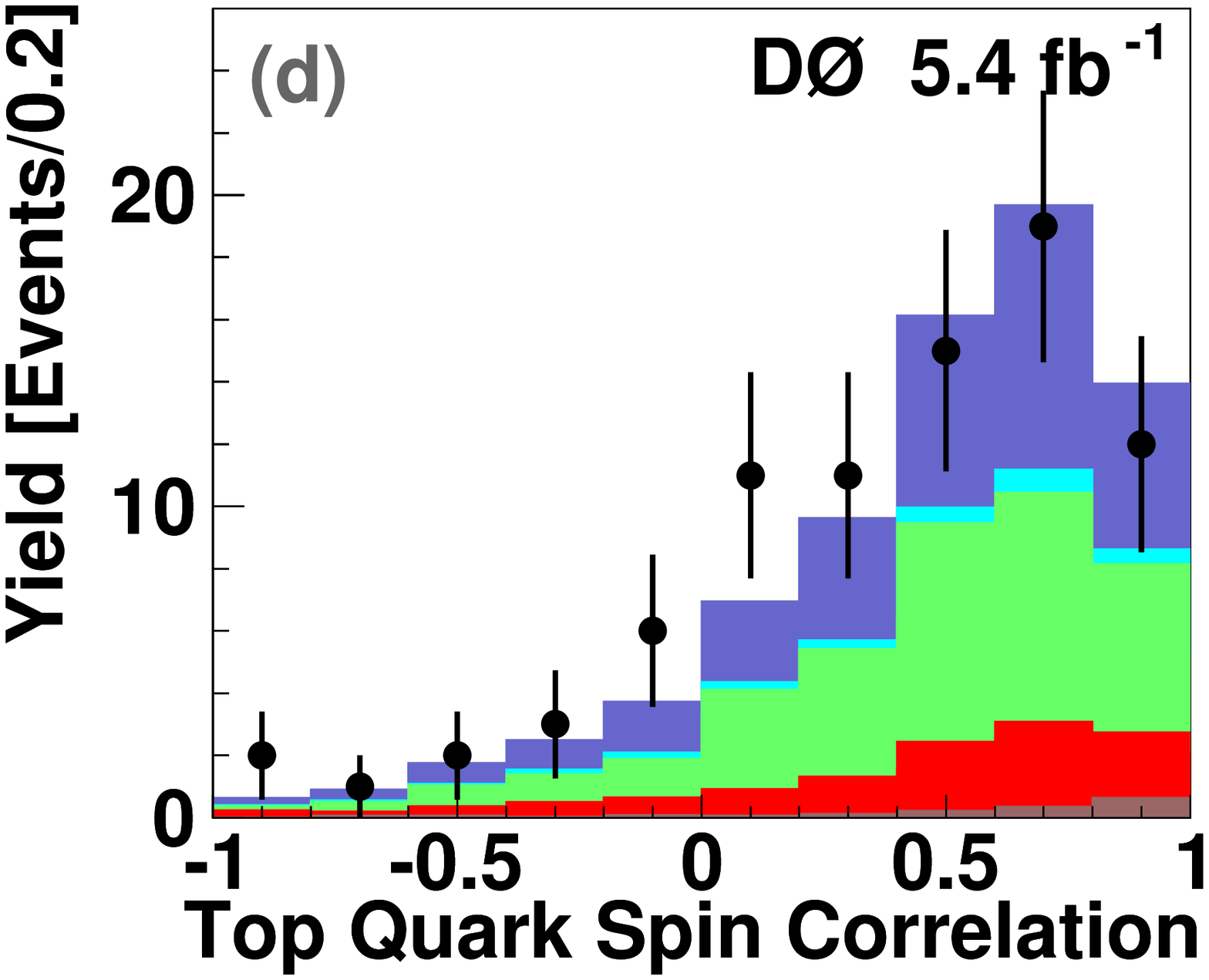}
\vspace{-0.2in}
\caption{(a) Total transverse energy, (b) reconstructed top quark mass, (c) lepton charge multiplied by the leading non-$b$-tagged quark jet pseudorapidity, and (d) $t$-channel top quark spin correlation (see text) for a sample selected with S:B~$>0.32$ using BNNComb discriminant. The $t$-channel and $s$-channel contributions have been normalized to their measured cross sections.}
\label{fig:distributions}
\end{figure}

The presence of the $t$-channel signal is visible in Fig.~\ref{fig:distributions}, which 
shows comparisons between the data, the background model, and the $t$-channel signal 
for a sample selected with S:B~$>0.32$ using the combined discriminant.
Four kinematic variables with large discriminating power for $t$-channel single top quark production
are shown:
$H_T$, the reconstructed top quark mass,
the lepton charge multiplied by the pseudorapidity of the leading non-$b$-tagged jet,
and the $t$-channel spin correlation in the optimal basis~\cite{Mahlon:1996pn,Cao:2005pq},
i.e., the cosine of the angle between the light quark jet and the lepton, both
in the reconstructed top quark rest frame.

The measured cross section depends on the assumed
mass of the top quark, $m_t$. The dependence is studied
by repeating the analysis on MC samples generated at
different values of $m_t$. Table~\ref{tab:BNNcombTMass} 
summarizes the measured cross section for different top quark masses. 

\begin{table}[!h!tbp]
\vspace{-0.1in}
\begin{center}
\caption{Variation of the measured cross section for $tqb$ and $tb$ single top quark production with
the top quark mass.}
\begin{ruledtabular}
\begin{tabular}{lccc}
$m_t$   & 170 GeV                & 172.5 GeV              & 175 GeV \\
\hline
$tqb$ & $2.80^{+0.57}_{-0.61}$ & $2.90^{+0.59}_{-0.59}$ & $2.53^{+0.58}_{-0.57}$ \\ 
$tb$ & $1.31^{+0.77}_{-0.74}$ & $0.98^{+0.62}_{-0.63}$ & $0.65^{+0.51}_{-0.50}$\\
\end{tabular}
\end{ruledtabular}
\vspace{-0.1in}
\label{tab:BNNcombTMass}
\vspace{-0.15in}
\end{center}
\end{table}


In summary, we have presented a model-independent measurement of the $t$-channel single top quark 
production cross section using $5.4\;\rm fb^{-1}$ of integrated luminosity 
recorded by the D0 experiment at the Fermilab Tevatron accelerator. 
For a top quark mass of $m_t = 172.5\;\rm GeV$, the measured cross section is
$\sigma({\ppbar}{\rargap}tqb+X) = 2.90 \pm 0.59\;\rm pb$. 
This is the most precise measurement to date, and is in good agreement with SM expectations.


%
We thank the staffs at Fermilab and collaborating institutions,
and acknowledge support from the
DOE and NSF (USA);
CEA and CNRS/IN2P3 (France);
FASI, Rosatom and RFBR (Russia);
CNPq, FAPERJ, FAPESP and FUNDUNESP (Brazil);
DAE and DST (India);
Colciencias (Colombia);
CONACyT (Mexico);
KRF and KOSEF (Korea);
CONICET and UBACyT (Argentina);
FOM (The Netherlands);
STFC and the Royal Society (United Kingdom);
MSMT and GACR (Czech Republic);
CRC Program and NSERC (Canada);
BMBF and DFG (Germany);
SFI (Ireland);
The Swedish Research Council (Sweden);
and
CAS and CNSF (China).

\vspace{-0.3in}


\end{document}